# Composite Consensus-Building Process: Permissible Meeting Analysis and Compromise Choice Exploration

Yasuhiro Asa[1][0000-0002-0349-6537], Takeshi Kato[2][0000-0002-6744-8606] and Ryuji Mine[1][0000-0002-0130-6752]

[1] Hitachi Kyoto University Laboratory, Hitachi Ltd., Kyoto, 606-8501, Japan
[2] Hitachi Kyoto University Laboratory, Kyoto University, Kyoto, 606-8501, Japan
yasuhiro.asa.mk@hitachi.com

**Abstract.** In solving today's social issues, it is necessary to determine solutions that are acceptable to all stakeholders and collaborate to apply them. The conventional technology of "permissive meeting analysis" derives a consensusable choice that falls within everyone's permissible range through mathematical analyses; however, it tends to be biased toward the majority in a group, making it difficult to reach a consensus when a conflict arises. To support consensus building (defined here as an acceptable compromise that not everyone rejects), we developed a composite consensus-building process. The developed process addresses this issue by combining permissible meeting analysis with a new "compromise choice-exploration" technology, which presents a consensusable choice that emphasizes fairness and equality among everyone when permissible meeting analysis fails to do so. When both permissible meeting analysis and compromise choice exploration do not arrive at a consensus, a facility is provided to create a sublated choice among those provided by them. The trial experimental results confirmed that permissive meeting analysis and compromise choice exploration are sufficiently useful for deriving consensusable choices. Furthermore, we found that compromise choice exploration is characterized by its ability to derive choices that control the balance between compromise and fairness. Our proposed composite consensus-building approach could be applied in a wide range of situations, from local issues in municipalities and communities to international issues such as environmental protection and human rights issues. It could also aid in developing digital democracy and platform cooperativism.

# 1 Introduction

Today's social issues are exacerbated because of social conflicts caused by diverse values, such as order formation in international conflicts, environmental conservation such as global warming countermeasures, countermeasures for pandemics such as Covid-19, and human rights issues such as well-being and gender equality. Otherwise expressed, social issues are deeply related to problems arising from conflicting opinions in groups, although there may be many opinions in such groups, and group decision-making methods are needed to solve these problems.

Typically, in the real world, voting via preference aggregation rules is widely used for group decision-making. Various types of preference aggregation rules exist, including simple majority voting, scoring rules such as the Borda count, and cumulative voting [1, 2]. However, these preference aggregation rules can cause cyclical preferences and can make consensus impossible (Condorcet's paradox) [3], and it has been proven that the conditions of fairness, Pareto efficiency (unanimity), completeness and transitivity of preference relations, independence of choices, and non-dictatorship, cannot be satisfied simultaneously (Arrow's impossibility theorem); therefore, it is not possible to make unique and fair decisions by voting [4].

Social choice theory is an academic field for social group decision-making that is desirable for all society members. One of the philosophies in this theory is "the greatest happiness of the greatest number (utilitarian principle)" proposed by the philosopher Bentham, which considers the relief of many poor people in a society of inequality as justice [5]. By contrast, the philosopher Rawls proposed that it is just to make the disparate society itself, which Bentham assumed equal and fair, and advocated "the free pursuit of happiness (the Liberty Principle)," "maximization of the benefits of the most disadvantaged (the Difference Principle)," and "fair equality of opportunity in duties and positions (the Fair Equality of Opportunity Principle)" [6]. Group decision-making based on Bentham's principle, like the voting described above, tends to bias the result toward the majority. Conversely, although Rawls' principles value diversity and fairness and respect for disadvantaged minorities, they are difficult to realize through voting.

Such group decision-making methods that emphasize individual freedom, equality, and fairness include discussion and deliberation. The social philosopher Habermas advocated the importance of group decision-making that emphasizes intersubjectivity through communicative acts in discussion [7]. In addition, the political scientist Fishkin has described the importance of decision-making based on changes in participants' opinions before and after deliberation through deliberative polling [8]. The political scientist Gutmann proposed deliberative democracy in which people listen to the opinions of others as a form of democracy through discussion [9], and the necessity and practicality of deliberative democracy in the field of education and urban development have been examined in Japan as well [10, 11].

The process of group decision-making through discussion and deliberation is important. Social psychologists Thibaut and Walker have shown that free discussion among participants can lead to satisfaction with the outcome and a sense of fairness [12], and psychologist Leventhal has described the factors in the process that promote a sense of fairness (procedural fairness criteria) [13]. The anthropologist Graeber has stated that original democracy is "a process of compromise and synthesis in which no one goes so far as to refuse to agree" [14]. Following Graeber's argument, "consensus" in this study is defined as an acceptable compromise that not everyone rejects. Specific methods for such a process include the spokes council in which spokes (representatives) are determined in units of affinity groups and discussions are held [15], and the consensus-building method advocated by the urban planner Susskind [16].

To support the process of group decision-making through discussions, several technologies with online opinion aggregation and chat functions have been proposed. For example, Decidim [17] has a full range of auxiliary components, such as questionnaires and blogs, vTaiwan [18] automatically groups participants' opinions, Loomio [19] and Liqlid [20] visualize opinions through pie charts and word clouds, and D-Agree [21] has functions for automatically structuring opinions and facilitation via artificial intelligence (AI). However, although these technologies are expected to be very effective in activating and organizing opinions, they do not have a support function when opinions conflict, and it is difficult to reach consensus participants, as Graeber advocates.

A mathematical framework for resolving conflicts of opinion is the graph model for conflict resolution (GMCR) [22-26]. The GMCR expresses the structure of conflicting opinions in a graphical model and performs mathematical analysis based on the preference order of decision makers for each opinion to derive a consensusable choice that takes rationality and efficiency into account. However, there is a problem that the number of states to be analyzed is as huge as $A^N$ (A: number of choices, N: number of decision makers; computational complexity order $O(k^n)$, exponential time). Therefore, permissible meeting analysis (PMA) has been proposed as a technology to avoid the problem, referring to the idea of GMCR [27, 28]. This method derives a consensusable choice that considers permissibility by performing a mathematical analysis based on ordering the decision makers' preferences for the choices, and their permissibility. The number of states to be analyzed is practical as it is the same as the number of choices A (computational complexity order $O(n)$, linear time). However, as PMA gives priority to the consensusable choice with the smallest total adjustment when adjusting everyone's permissible range, it is easy to derive a consensusable choice that is less burdensome for the majority, and it may be difficult to reach a consensus, including the minority group.

Based on the above, our objective was to provide a new consensus-building support technology to reach consensus and a new consensus-building process for using this technology. Specifically, in addition to the conventional technology PMA, we provide a new technology compromise choice exploration (CCE). A composite consensus-building process that combines the two consensusable choices derived from both supports consensus. As mentioned, the problem is that PMA alone tends to present a consensusable choice that is biased toward the majority. Therefore, we provide a technology for presenting a consensusable choice that emphasizes fairness, where compromise among all is equalized by CCE when PMA fails to promote consensus. The new composite consensus-building process combines PMA and CCE to enable a consensus that falls within everyone's permissibility and range of compromise while emphasizing fairness. These technologies and processes expand the research streams of Rawls' principle of difference, the deliberative democracy of Gutmann et al., Leventhal's procedural justice, and Graeber's compromise and synthesis, and they put forth new research directions in terms of developing conventional opinion exchange tools and mathematical analysis models.

The remainder of this paper is organized as follows: In the "Related Works" section, we describe the related works and issues of the online discussion platform for consensus building and the mathematical modeling technology for resolving conflicts of opinion.

The "Methods" section describes the conventional PMA, our newly proposed CCE, and a new composite consensus-building process that combines them. In the "Results" section, we present the outcomes of testing our technology and process on the exemplar problem, and in the "Discussion" section, we discuss the effectiveness of our technology and process based on the "Results" and the remaining issues. In the last section, we present our "Conclusions."

## 2    Related Works

### 2.1    Discussion Support Technology

Among the operational technologies that assist discussion, there are online platforms that support multiple participants. Climate CoLab [29] is for creating solutions to climate change problems. It is a contest-style platform that invites ideas for solutions to problems and aggregates a variety of opinions. Since its inception in 2009, it has regularly held online conferences that, to date, have attracted more than 120,000 participants [30]. However, Climate CoLab's purpose is to propose diverse solutions, not to build consensus.

Decidim [17] is a leading online discussion platform for democratic self-organization. It is open source and widely used with 320,000 users and more than 160 projects. It is characterized by its rich functionality for discussions, such as surveys, meetings, proposals, budgets, debates, and blogs. However, it does not have a facilitation support function for discussions, and the activation and convergence of discussions depend on the participants.

D-Agree [21] is an online discussion platform similar to Decidim, which has a function to automatically perform facilitation by AI agents on behalf of humans to support discussion activation. This automatic facilitation function is realized by automatically extracting issue-based information system (IBIS) structures [31] from participants' opinions and automatically generates questions for participants using AI. However, it does not yet have a facilitation function to converge opinions.

Although it is impossible to list all the related works, there are several online discussion platforms such as Loomio [19], which visualizes the opinions for and against, vTaiwan [18], which automatically groups the participants' opinions for and against, and Liqlid [20], which visualizes the focus of discussion based on words that occur frequently in the expressed opinions.

Conventional technologies can effectively activate online discussions and organize opinions. However, they are not equipped with facilitation support functions that resolve conflicts of opinion. Therefore, we aimed to provide a technology for deriving a consensusable choice to resolve conflicts of opinion.

### 2.2    Conflict Resolution Technology

Other related work on conflict resolution technology have supported the process of deciding on a consensus choice among multiple choices. Group Navigator [32] determines a unique consensus choice by weighting each option by the participants based on a set of evaluation criteria. Then, discussions are held to negotiate the weighting values until everyone is satisfied with the consensus choice. However, it is highly probable that it will derive a consensusable choice biased toward the majority, resulting in a disregard for the minority group. Additionally, to calculate the weights, pairwise comparisons must be made for all combinations of choices and evaluation criteria, and recalculation is necessary when new choices or evaluation criteria are found during the discussion.

GMCR is a mathematical analysis method for resolving conflicts of opinion using a graph model developed by Hipel and Kilgour et al [22-26]. Specifically, the graph model is constructed by expressing the state of conflicting opinions for "combinations of choices that all participants can choose (states)," "preference order for the states of each participant," and "state transitions of each participant." Then, using two methods, rational analysis, which seeks a stable state (a state in which participants stop transitioning) based on action criteria (the number of anticipated state transitions), and efficiency analysis, which seeks a stable state based on the order of preference, the stable state that will eventually be settled is derived, respectively, and facilitation is conducted to reach this state. However, it cannot manage cases in which consensus cannot be reached among all in a stable state. In addition, the number of states $A^N$ (A: number of choices, N: number of decision makers) subject to preference order is huge, rendering it impractical.

PMA is a consensus-building support technology that can avoid GMCR's, practical issues while referring to the GMCR concept [27, 28]. Based on the participants' preference order for the choices and their permissible ranges, it derives a consensusable choice acceptable to all and the currently permissible range of conditions. Preference order targets only the number of choices, which is significantly less computationally intensive than GMCR. Facilitation is then performed based on this consensusable choice. However, this does not always lead to consensus. In addition, the consensusable choice tends to be biased toward majority importance because the option with the smallest possible change from the most preferred preference order is given priority.

Although it is practically impossible to list all related previous works, there are several studies that have expanded on the GMCR, including Xu et al. [33], who conducted a GMCR stability analysis using the preference order of decision makers based on their relationships, and Shahbaznezhadfard et al. [34], who studied the relationship between dynamic changes in GMCR parameters and conducted a stability analysis. Thus, in a related work, it was possible to derive a consensusable choice in a conflict of opinions by using the participants' preference order and a mathematical model. However, such consensusable choices tend to be biased toward the majority, and that consensus can possibly not be reached among all participants, including the minority. Therefore, we aimed to provide a new technology to derive a consensusable choice that emphasizes fairness when using PMA alone does not lead to consensus.

# 3 Methods

## 3.1 Permissible Meeting Analysis

First, we explain the conventional conflict resolution technology, PMA, as our composite consensus-building process employs PMA as one of its functions.

In a discussion, let $M = \{1,2,\cdots,m\}$ be the set of participants, $X = \{x_1, x_2, \cdots, x_n\}$ be the set of choices, and $\succsim_i = \{x_{i1}, x_{i2}, \cdots, x_{in}\}$ be the preference order for the choices of participant $i \in M$. Let $x_{ij} \in X$, $j$ be the preference order, $max\, P_i = \{x_{i1}, x_{i2}, \cdots, x_{ik_i} | k_i \leq n\}$ be the set of acceptable choices for participant $i$, and $max P_i^l = \{x_{i1}, x_{i2}, \cdots, x_{ik_i}, \cdots, x_{i(k_i+l)} | (k_i + l) \leq n\}$ be the set of acceptable choices for $max\, P_i$ extended by $l$.

The algorithm for PMA is shown in Fig. 1. In PMA, consensusable choices are those that exist within the permissible range of all participants. Therefore, in Step 1, the product set $U_o$ of acceptable choices for participant $i$ ($= 1,2,\cdots,m$) is derived by Eq. 1, and whether $U_o = \emptyset$ is determined. If $U_o = \emptyset$ does not hold, then the process moves to Step 2, outputs the set of alternatives $U_o$ derived in Eq. 1 as the consensusable choice $X_o$, and terminates.

If $U_o = \emptyset$ in Step 1, the permissible range $l$ of each participant $i$ is expanded individually in the loop of Steps 3, 4, and 6 to derive the union set $U_l$ of the permissible ranges of participant $i$ ($= 1,2,\cdots,m$), respectively, as shown in Eq. 2, and determine whether $U_l = \emptyset$. If $U_l = \emptyset$ is no longer the case in Step 4, then in Step 5, the consensusable choice is the set of choices in the product set $U_l$ that has the smallest permissible range of $l$, i.e., the set $X_l$ of choices that is the smallest in Eq. 3.

$$U_o = \cap_{i=1}^{m} max P_i \tag{1}$$

$$U_l = \cap_{i=1}^{m} max P_i^l \quad (l = 0, \cdots, n - k_i) \tag{2}$$

$$X_l = \underset{U_l}{\operatorname{argmin}} \sum_{i=1}^{m} l \tag{3}$$

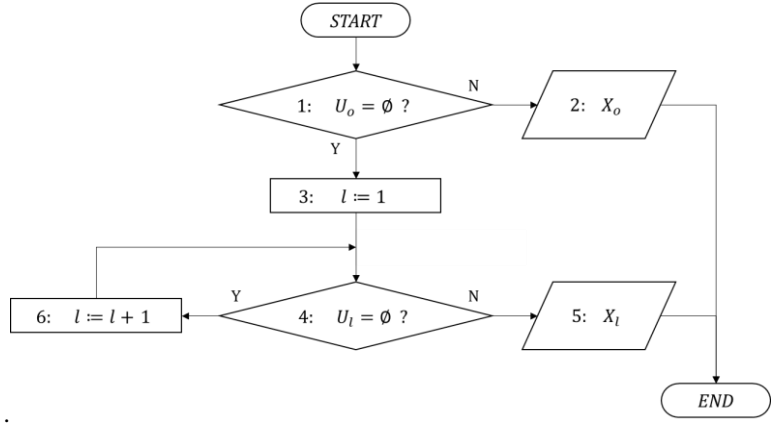

**Fig. 1.** Flowchart of permissible meeting analysis

If the set $X_o$ in Step 2 or the set $X_l$ in Step 5 contains more than one consensusable choice, all of them shall be considered as the result of the derivation from PMA. Note that the loop process in Steps 3, 4, and 6 does not fall into an infinite loop because there is always a permissible range $l$ such that $U_l \neq \emptyset$.

For example, as shown in the second line of Table 1, if the set of participants is $M = \{1,2,3\}$, the set of choices is $X = \{a, b, c\}$, the preference order for the choices of each participant is $\succsim_1 = \{a, b, c\}$, $\succsim_2 = \{b, c, a\}$, $\succsim_3 = \{c, a, b\}$, and the permissible range of each participant is $max\ P_1 = \{a, b\}$, $max\ P_2 = \{b\}$, $max\ P_3 = \{c, a\}$, then in Step 1, Eq. 1 gives $U_o = \emptyset$; thus, no consensusable choice exists within the permissible range of all participants. Therefore, in Step 4, the product set $U_l$ is obtained when the permissible range of each participant is gradually increased by Eq. 2, as shown below in the third line of Table 1. Then, in Step 5, Eq. 3 derives a consensusable choice $X_l = \{b\}$ that minimizes $\sum_{i=1}^{3} l$ from this product set. In other words, if participant 3 extends the permissible range from $\{c, a\}$ to $\{c, a, b\}$, the consensusable choice $\{b\}$ becomes an option within the permissible range of all participants. Although only a few examples are listed in Table 1, it suggests that participants 1 and 2 do not need a wider permissible range and that PMA tends to be biased in favor of the majority. This is also discussed in the "Results" section through Table 7.

**Table 1.** Relationship between permissible range and consensusable choices

| | Permissible Range | | | Permissible Sets | Permissible Sets | Permissible Sets | Consensusable Choices |
|---|---|---|---|---|---|---|---|
| $k_1 + l$ | $k_2 + l$ | $k_3 + l$ | $\sum_{i=1}^{3} l$ | $\max P_1^l$ | $\max P_2^l$ | $\max P_3^l$ | $U_l = \bigcap_{i=1}^{3} \max P_i^l$ |
| 2 | 1 | 2 | 0 | $\{a, b\}$ | $\{b\}$ | $\{c, a\}$ | $\emptyset$ |
| 3 | 1 | 2 | 1 | $\{a, b, c\}$ | $\{b\}$ | $\{c, a\}$ | $\emptyset$ |
| 2 | 2 | 2 | 1 | $\{a, b\}$ | $\{b, c\}$ | $\{c, a\}$ | $\emptyset$ |
| 2 | 1 | 3 | 1 | $\{a, b\}$ | $\{b\}$ | $\{c, a, b\}$ | $\{b\}$ |
| 3 | 2 | 2 | 2 | $\{a, b, c\}$ | $\{b, c\}$ | $\{c, a\}$ | $\{c\}$ |
| 2 | 2 | 3 | 2 | $\{a, b\}$ | $\{b, c\}$ | $\{c, a, b\}$ | $\{b\}$ |
| 2 | 3 | 2 | 2 | $\{a, b\}$ | $\{b, c, a\}$ | $\{c, a\}$ | $\{a\}$ |
| 3 | 3 | 2 | 3 | $\{a, b, c\}$ | $\{b, c, a\}$ | $\{c, a\}$ | $\{c, a\}$ |
| 2 | 3 | 3 | 3 | $\{a, b\}$ | $\{b, c, a\}$ | $\{c, a, b\}$ | $\{a, b\}$ |
| 3 | 3 | 3 | 4 | $\{a, b, c\}$ | $\{b, c, a\}$ | $\{c, a, b\}$ | $\{a, b, c\}$ |

### 3.2 Compromise Choice Exploration

Based on the PMA described in the previous section, we propose our new consensusable choice derivation technology, CCE. In PMA, the participants' act of compromise is to expand the permissibility range for the preferential order of choices, and as a consensusable choice is derived that minimizes the overall act of compromise, the compromise is likely to be biased toward the minority rather than the majority, resulting in a lack of fairness. Therefore, in CCE, the participants' act of replacement ordering for their preference order is regarded as a compromise act by the participants, and a common preference order that makes the number of replacement operations by each participant as equal as possible is regarded as a consensusable choice. This will lead to a highly fair consensusable choice.

The algorithm for CCE is shown in Fig. 2. CCE searches for a preference order among the list $(\succsim)_{all}$ of all preference orders for choice $X$, such that the number of order replacements from the initial preference order $\succsim_i$ for each participant $i$ is equal. First, Step 1 generates $n!$ lists $(\succsim)_{all}$. Next, to calculate the number of replacements between the preference order $(\succsim_i)_{i \in M}$ of each participant $i$ and each preference order of $(\succsim)_{all}$ for this list, in Step 3, as in Eq. 4, each element of the preference order of participant $i$ is replaced by a number in ascending order according to the replacement rule $Rule_i$ in ascending numerical order as shown in Eq. 4.

$$Rule_i = \{x_{i1} \rightarrow 1, x_{i2} \rightarrow 2, \cdots, x_{in} \rightarrow n\} \tag{4}$$

In Step 5, we replace each preference order in the preference order list $(\succsim)_{all}$ with the same rule $Rule_i$ as in Eq. 4, as in Eq. 5. where $j = 1, \cdots, n!$.

$$\succsim'_j = \left\{ x_{j1} \rightarrow x'_{j1}, x_{j2} \rightarrow x'_{j2}, \cdots, x_{jn} \rightarrow x'_{jn} \right\} \tag{5}$$

In Step 6, each preference order in the preference order list $(\succsim)'_{all}$ replaced by Eq. 5 is sorted in ascending order as shown in Eq. 6, and the number of sortings $r_{js}$ at that time is calculated according to Eq. 7.

$$\succsim'_{js} = Sort\left( x'_{j1}, x'_{j2}, \cdots, x'_{jn} \right) \tag{6}$$

$$r_{js} = SortCount\left( x'_{j1}, x'_{j2}, \cdots, x'_{jn} \right) \tag{7}$$

In the loops of Steps 7 and 8, and Steps 9 and 10, the process in Steps 3, 5, and 6 is performed for all participants $i$ and all choices $j$.

In Step 11, based on the number of replacements for each list in the preference order list $(\succsim)'_{all}$ obtained from each participant's preference order $\succsim_i$, we search for the one among $(\succsim)'_{all}$ for which the number of replacements for all participants is equal. Specifically, based on Eqs. 8 and 9, the average $\mu$ and *Standard deviation* $\sigma$ of the number of replacements $r_i$ $(i = 1, \cdots, m)$ for each participant $i$ for each list of $(\succsim)'_{all}$ are derived and the *Score* is calculated.

$$\sigma = \sqrt{\frac{1}{m} \sum_{i=1}^{m} (r_i - \mu)^2} \tag{8}$$

$$Score = \mu + \sigma \tag{9}$$

In Step 12, the list of preference order that minimizes the *Score* in Eq. 10 is selected. The average $\mu$ is smaller the less the degree of compromise from the initial preference order. The *Standard deviation* $\sigma$ is smaller for lists where all participants have the same number of replacements (when the value is 0, all participants have the same number of replacements). Therefore, Eq. 9 has a smaller *Score* when the degree of compromise by each participant is as small as possible and as fair as possible, and Eq. 10 derives the option with the smallest *Score* as the most consensusable choice.

$$X_s = \underset{(\succsim)'_{all}}{\operatorname{argmin}} Score \tag{10}$$

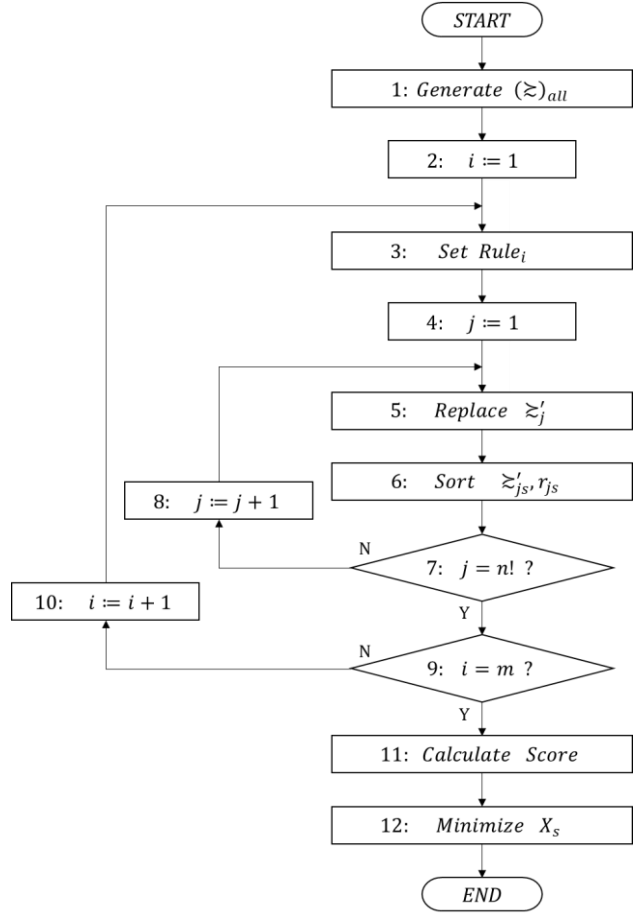

**Fig. 2.** Flowchart of compromise choice exploration

For example, as shown in Fig. 3, if the set of participants is $M = \{1,2,3\}$, the set of choices is $X = \{a, b, c, d\}$, and the preference order for the choices of each participant is $\succsim_1 = \{a, b, c, d\}$, $\succsim_2 = \{d, a, c, b\}$, $\succsim_3 = \{c, d, a, b\}$, then the list $(\succsim)_{all}$ of all preference order for the choices $X$ is $4! (= 24)$ ways. Here, the edges in Fig. 3 correspond to the replacement operations. Using Eqs. 4–7 according to Steps 1–10, the number of replacements between each participant's preference order and each list in the preference order list $(\succsim)_{all}$ is obtained, and the results are shown in Table 2. Then, *Score* is obtained for each list using Eqs. 8–10 according to Step 11 and arranged in ascending order to obtain the result in Table 3. In Step 12, the preference order $X_s = \{a, c, d, b\}$ with the smallest *Score* is derived, and the first-ranked $\{a\}$ is derived as the consensusable choice.

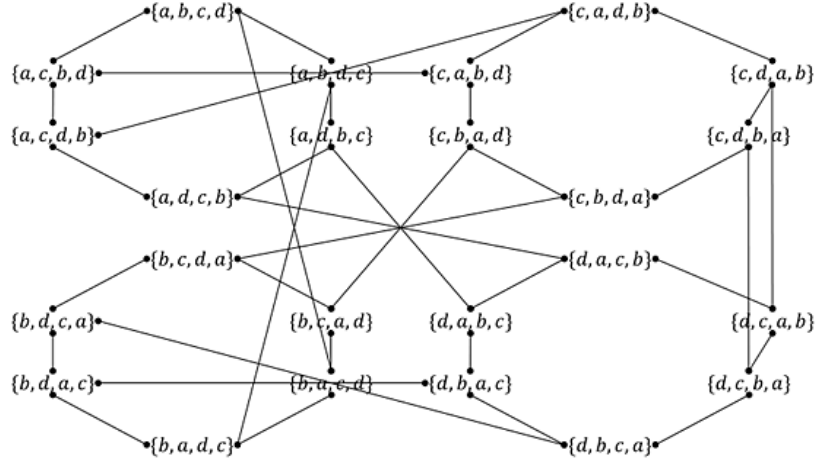

**Fig. 3.** Preference order switching operation (Choices: $a, b, c, d$)

**Table 2.** Number of replacements between each participant's preference order and $(\succsim)_{all}$

| $(\succsim)_{all}$ | $Rule_1 = \{a \to 1, b \to 2, c \to 3, d \to 4\}$ | | $Rule_2 = \{d \to 1, a \to 2, c \to 3, b \to 4\}$ | | $Rule_3 = \{c \to 1, d \to 2, a \to 3, b \to 4\}$ | |
|---|---|---|---|---|---|---|
| | $\succsim'_1$ | $r_{js}$ | $\succsim'_2$ | $r_{js}$ | $\succsim'_3$ | $r_{js}$ |
| $b,c,a,d$ | 2,3,1,4 | 2 | 4,3,2,1 | 6 | 4,1,3,2 | 4 |
| $a,c,b,d$ | 1,3,2,4 | 1 | 2,3,4,1 | 3 | 3,1,4,2 | 3 |
| $a,b,d,c$ | 1,2,4,3 | 1 | 2,4,1,3 | 3 | 3,4,2,1 | 5 |
| $\vdots$ | $\vdots$ | $\vdots$ | $\vdots$ | $\vdots$ | $\vdots$ | $\vdots$ |

**Table 3.** Compromise exploration score

| $(\succsim)_{all}$ | $r_{js}$ for $\succsim'_1$ | $r_{js}$ for $\succsim'_2$ | $r_{js}$ for $\succsim'_3$ | Average "$\mu$" | Standard deviation "$\sigma$" | Score "$\mu + \sigma$" |
|---|---|---|---|---|---|---|
| $a,c,d,b$ | 2 | 2 | 2 | 2 | 0 | 2 |
| $b,d,c,a$ | 4 | 4 | 4 | 4 | 0 | 4 |
| $c,a,d,b$ | 3 | 3 | 1 | 2.333 | 0.943 | 3.276 |
| $\vdots$ | $\vdots$ | $\vdots$ | $\vdots$ | $\vdots$ | $\vdots$ | $\vdots$ |

Note that because the number of $(\succsim)_{all}$ to be fully searched in CCE is $n!$ (the computational complexity order is $O(n!)$, factorial time), the computational load increases as the number of choices increases. However, the number of choices managed in actual consensus building is approximately 10 at most, and it is considered that a full search is not problematic for practical use. Although we used a full search here, a more efficient algorithm that gradually expands the replacement network from the participant's initial preference order and searches for the option with the smallest *Score* could be considered instead.

### 3.3 Composite Consensus-building Process

In addition to the PMA and CCE described so far, we propose a new composite consensus-building process that combines the PMA and CCE with sublated choice creation (SCC), which in turn creates a new sublated choice from multiple choices derived from the former two. Figure 4 depicts the flow of the consensus-building process for the problem of "deciding on the option that all participants agree from multiple choices."

First, in Step 1, PMA is performed based on the algorithm described in Section 3.1, and based on the results, in Step 2, the participants discuss whether consensus is possible. In this discussion, if the same choices exist within everyone's permissible range from the beginning and consensus is reached, then the discussion may be terminated. If the same choice does not exist, the consensusable choice derived in the Step 1 PMA and the permissible range conditions at that time are presented to the participants, and if consensus is reached among all participants, the process is terminated.

If no consensus is reached among all participants in Step 2, they proceed to Step 3 and perform CCE based on the algorithm described in Section 3.2. Because the nature of the consensusable choices derived by PMA tends to be biased in favor of the majority, if consensus is not reached in Step 2, attempted fairness to the participants is considered to be the cause. The consensusable choice derived in CCE is the fairer choice and is more likely to achieve consensus. Therefore, in Step 4, the consensusable choice derived in CCE is presented to the participants to discuss the possibility of consensus. If consensus can be reached in this discussion, the process will be terminated.

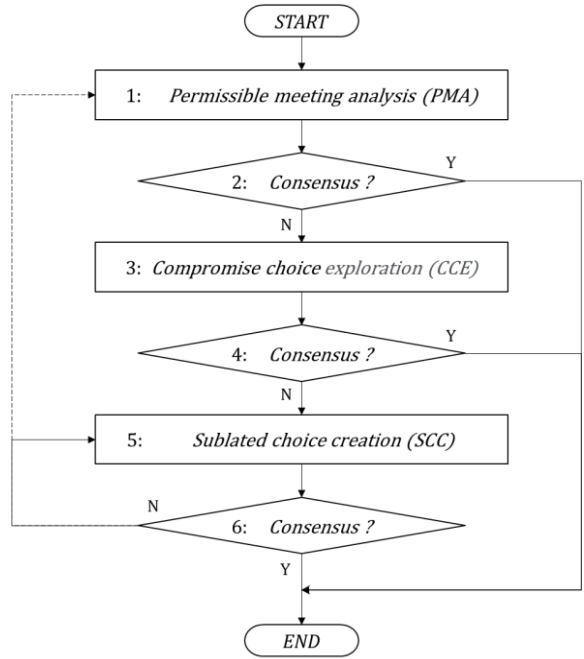

**Fig. 4.** Proposed composite consensus-building process

If no consensus is reached in Step 4, proceed to Step 5 to perform a new SCC based on the results of Step 1 (PMA) and Step 3 (CCE). Specifically, the contents of the choices derived by PMA that are likely to fall within everyone's permissibility range and the contents of the choices derived by CCE with high-rank preference order based on fairness are presented to the participants, and a new choice (sublated choice) is created by combining these contents. Then, in Step 6, consensusability will be discussed based on this sublated choice. If consensus is reached in this discussion and a consensus choice is determined, the process is terminated.

If Step 6 still does not produce a consensus, return to Step 5, increase the number of consensusable choices derived from PMA and CCE, and recreate a new sublated choice. Specifically, PMA uses the choice with the smallest Eq. 3 and the second-smallest choice, and CCE uses the choice with the smallest Eq. 9 and the second-smallest choice, thereby creating a new sublated choice that combines them. Then, based on the sublated choice, the discussion is held again in Step 6.

Here, if multiple recreated sublated choices are created, it is possible to return to the Step 1 process as PMA and CCE are possible depending on the preference order and permissible range of the participants for these multiple sublated choices.

Thus, in our newly proposed composite consensus-building process, aiming for consensus, we first use PMA for a discussion based on a consensusable choice that falls within everyone's permissible range, and if consensus cannot be reached, we use CCE for a discussion based on a consensusable choice that considers fairness. Then, if consensus cannot be reached using either, a sublated choice is created by synthesizing the consensusable choices of PMA and CCE and is discussed. Conventional processes allow no effective means of supporting participants to reach a consensus when there are conflicts of opinion, but this process can support facilitation by presenting an effective consensusable choice.

## 4      Results

### 4.1      Trial Setup

To evaluate our proposed consensus-building support technology, we conducted trial experiments on PMA and CCE under the theme presented in Table 4. As there is no standard benchmark dataset on consensus building, we developed our own dataset in this study. The theme was "How to handle nuclear power generation in Japan for the future," and there were five participants (persons A to E: four males and one female; age ranging from 30s to 60s) and seven choices, namely (1) zero nuclear power plants by 2030; (2) nationalize and decommission nuclear power plants; (3) no new nuclear power plants, but restarting nuclear power plants is possible on the conditions of safety and local consent; (4) no new nuclear power plants, but restarting nuclear power plants is possible until alternative power generation methods are established; (5) restart nuclear power plants to decommission them and promote the development of next-generation nuclear power plants; (6) nuclear power plants can be operated with emphasis on safety; and (7) proactively utilize nuclear power plants.

**Table 4.** Consensus building: Theme and choices

| Theme | Discussion | Choices |
|---|---|---|
| Nuclear Power Generation | How should nuclear power be managed in Japan's future energy policy? | (1) Zero nuclear power plants by 2030. |
| | | (2) Nationalize and decommission nuclear power plants. |
| | | (3) No new nuclear power plants; but restarting nuclear power plants is possible on the condition of safety and local consent. |
| | | (4) No new nuclear power plants; but restarting nuclear power plants is possible until alternative power generation methods are established. |
| | | (5) Restart nuclear power plants intending to decommission them and promote the development of next-generation nuclear power plants. |
| | | (6) Nuclear power plants can be operated with emphasis on safety. |
| | | (7) Proactively utilize nuclear power plants. |

The discussion was hosted on the D-Agree online platform, and the chat function was used to collect each participant's opinion on the theme, as well as the preference order for the choices and permissible ranges. As mentioned, D-Agree has an automatic facilitation function, which means that when an opinion is posted by a participant, an intervention is made in response to that opinion to encourage others to contribute, thereby stimulating discussion.

### 4.2 Trial Results

Table 5 shows the results of obtaining the preference order for the choices and permissible ranges for the seven choices of the five participants for the theme in Table 4.

**Table 5.** Preference order and permissible range (white: permitting, gray: not permitting)

| | "A" | "B" | "C" | "D" | "E" |
|---|---|---|---|---|---|
| Rank 1 | (5) | (4) | (4) | (5) | (4) |
| Rank 2 | (4) | (3) | (3) | (4) | (2) |
| Rank 3 | (3) | (2) | (2) | (3) | (1) |
| Rank 4 | (2) | (6) | (1) | (2) | (3) |
| Rank 5 | (1) | (1) | (5) | (1) | (5) |
| Rank 6 | (6) | (7) | (6) | (6) | (6) |
| Rank 7 | (7) | (5) | (7) | (7) | (7) |

PMA was conducted for the results presented in Table 5 according to the method described in Section 3.1. The resulting consensusable choice was (4) "no new nuclear power plants but restarting nuclear power plants is possible until alternative power generation methods are established," which was within the permissible range for all the participants.

Table 6 shows the CCE results according to the method described in Section 3.2. By rearranging the preference order based on the *Score*, which is the sum of the average $\mu$ and the *Standard deviation* $\sigma$, the preference order {(4),(3),(2),(5),(1),(6),(7)} with the smallest *Score* of 3.78 and the first-ranked choice (4) "no new nuclear power plants, but restarting nuclear power plants is possible until alternative power generation methods are established" are derived as the consensusable choices. *Score* is the sum of $\mu$, which indicates the number of replacements (compromises), and $\sigma$, which indicates fairness. In Table 6, instead of using Eq. 9, the preference order {(4),(3),(2),(1),(5),(6),(7)} can be derived by selecting the choice of 2.6, which is the minimum value of the number of replacements $\mu$, and the preference order {(4),(2),(5),(3),(6),(1),(7)} can be derived by selecting the choice of 0.4, which is the minimum value of fairness $\sigma$. The right-hand side of Eq. 9, which weighs more heavily on the number of replacements or on fairness, is discussed below in the "Discussion" section.

**Table 6.** Results of the compromise choice exploration

| $(\succeq)_{all}$ | $r_{js}$ for $\succeq'_A$ | $r_{js}$ for $\succeq'_B$ | $r_{js}$ for $\succeq'_C$ | $r_{js}$ for $\succeq'_D$ | $r_{js}$ for $\succeq'_E$ | Average "μ" | Standard deviation "σ" | Score "μ + σ" |
|---|---|---|---|---|---|---|---|---|
| {(4),(3),(2),(5),(1),(6),(7)} | 3 | 4 | 1 | 3 | 3 | 2.8 | 0.98 | 3.78 |
| {(4),(3),(2),(1),(5),(6),(7)} | 4 | 3 | 0 | 4 | 2 | 2.6 | 1.5 | 4.1 |
| {(4),(3),(2),(5),(6),(1),(7)} | 4 | 3 | 2 | 4 | 4 | 3.4 | 0.8 | 4.2 |
| {(4),(3),(5),(2),(1),(6),(7)} | 2 | 5 | 2 | 2 | 4 | 3 | 1.26 | 4.26 |
| {(4),(3),(5),(2),(6),(1),(7)} | 3 | 4 | 3 | 3 | 5 | 3.6 | 0.8 | 4.4 |
| {(4),(2),(5),(3),(6),(1),(7)} | 4 | 5 | 4 | 4 | 4 | 4.2 | 0.4 | 4.6 |
| {(3),(4),(2),(5),(1),(6),(7)} | 4 | 5 | 2 | 4 | 4 | 3.8 | 0.98 | 4.78 |
| ⋮ | ⋮ | ⋮ | ⋮ | ⋮ | ⋮ | ⋮ | ⋮ | ⋮ |

As mentioned earlier, within the participants' preference order and permissible range shown in Table 5, choice (4) was present in everyone's permissible range from the beginning, choice (4) chosen by PMA coincided with choice (4) chosen by CCE, and there was no conflict of opinion state. Therefore, to evaluate a case where there is a conflict of opinions, we assume, for example, as shown in Table 7, that there is no consensus choice within the permissible range of all the participants. This corresponds to the process from Step 1 to Steps 3 and 4 in Fig. 1.

**Table 7.** Preference order and permissible range (white: permitting, gray: not permitting)

| | "A" | "B" | "C" | "D" | "E" |
|---|---|---|---|---|---|
| Rank 1 | (5) | (4) | (7) | (5) | (6) |
| Rank 2 | (4) | (3) | (6) | (4) | (7) |
| Rank 3 | (3) | (2) | (2) | (3) | (1) |
| Rank 4 | (2) | (6) | (1) | (2) | (5) |
| Rank 5 | (1) | (1) | (4) | (1) | (3) |
| Rank 6 | (6) | (7) | (3) | (6) | (4) |
| Rank 7 | (7) | (5) | (5) | (7) | (2) |

Table 8 shows the results of PMA for the assumptions in Table 7. In this case, choice (1) "zero nuclear power plants by 2030," which has the smallest value of 2 for $\sum_{i=1}^{5} l$, which represents the degree of widening of the overall permissible range, is derived as the consensusable choice, and the permissible range conditions at that time are to widen the permissible ranges of two persons, B and D, from 4 to 5. Although the examples in Table 7 are few, this suggests that persons A, C, and E do not need a wider permissible range and that PMA tends to be biased in favor of the majority.

Note that Table 8 shows that there are three cases where the value of $\sum_{i=1}^{5} l$ is 3, and their consensusable choices are different from (2), (4), and (3). This suggests that although one consensusable choice could be selected in the hypothetical example in Table 7, in some situations, there are cases in which the value of $\sum_{i=1}^{5} l$ alone does not narrow down to a single consensusable choice. This is discussed below in the "Discussion" section.

**Table 8.** Results of the permissible meeting analysis for Table 7

| | | | | | | | | | | | |
|---|---|---|---|---|---|---|---|---|---|---|---|
| *Permissible Range* | | | | | | | | | | | |
| $k_B + l$ | $k_B + l$ | $k_C + l$ | $k_D + l$ | $k_E + l$ | $\sum_{i=1}^{5} l$ | $max\, P_A^l$ | $max\, P_B^l$ | $max\, P_C^l$ | $max\, P_D^l$ | $max\, P_E^l$ | $\bigcap_{i=1}^{5} max P_i^l$ |
| 5 | 5 | 4 | 5 | 4 | 2 | (5)(4)(3)(2)(1) | (4)(3)(2)(6)(1) | (7)(6)(2)(1) | (5)(4)(3)(2)(1) | (6)(7)(1)(5) | (1) |
| 5 | 4 | 4 | 4 | 7 | 3 | (5)(4)(3)(2)(1) | (4)(3)(2)(6) | (7)(6)(2)(1) | (5)(4)(3)(2) | (6)(7)(1)(5)(3)(4)(2) | (2) |
| 5 | 4 | 5 | 4 | 6 | 3 | (5)(4)(3)(2)(1) | (4)(3)(2)(6) | (7)(6)(2)(1)(4) | (5)(4)(3)(2) | (6)(7)(1)(5)(3)(4) | (4) |
| 5 | 4 | 6 | 4 | 5 | 3 | (5)(4)(3)(2)(1) | (4)(3)(2)(6) | (7)(6)(2)(1)(4)(3) | (5)(4)(3)(2) | (6)(7)(1)(5)(3) | (3) |
| ⋮ | ⋮ | ⋮ | ⋮ | ⋮ | ⋮ | ⋮ | ⋮ | ⋮ | ⋮ | ⋮ | ⋮ |

Table 9 shows the results of CCE performed on Table 7. The preference order {(4),(6),(7),(5),(3),(2),(1)} that minimizes *Score* and the first-ranked choice (4) "no new nuclear power plants, but nuclear power plants can be restarted until alternative power generation methods are established" are derived as consensusable choices. However, Table 9 shows several cases in which *Score* and *Standard deviation* $\sigma$ have the same value. This suggests that although one consensusable choice could be selected in the assumed example in Table 7, in some cases, the *Score* or *Standard deviation $\sigma$* alone may be insufficient to narrow the consensus to one consensusable choice. This is discussed below in the "Discussion" section.

**Table 9.** Results of the compromise choice exploration for Table 7

| $(\succsim)_{all}$ | $r_{js}$ for $\succsim'_A$ | $r_{js}$ for $\succsim'_B$ | $r_{js}$ for $\succsim'_C$ | $r_{js}$ for $\succsim'_D$ | $r_{js}$ for $\succsim'_E$ | Average "$\mu$" | Standard deviation "$\sigma$" | Score "$\mu + \sigma$" |
|---|---|---|---|---|---|---|---|---|
| {(4),(6),(7),(5), (3),(2),(1)} | 9 | 8 | 10 | 9 | 8 | 8.8 | 0.748 | 9.548 |
| {(4),(6),(7),(5), (2),(3),(1)} | 10 | 9 | 9 | 10 | 9 | 9.4 | 0.49 | 9.89 |
| {(4),(7),(6),(5), (3),(2),(1)} | 10 | 9 | 9 | 10 | 9 | 9.4 | 0.49 | 9.89 |
| {(4),(6),(5),(2), (7),(1),(3)} | 9 | 10 | 10 | 9 | 10 | 9.6 | 0.49 | 10.09 |
| {(6),(4),(5),(2), (7),(3),(1)} | 9 | 10 | 10 | 9 | 10 | 9.6 | 0.49 | 10.09 |
| {(6),(4),(5),(2), (1),(7),(3)} | 9 | 10 | 10 | 9 | 10 | 9.6 | 0.49 | 10.09 |
| {(4),(7),(5),(6), (3),(2),(1)} | 9 | 10 | 10 | 9 | 10 | 9.6 | 0.49 | 10.09 |
| ⋮ | ⋮ | ⋮ | ⋮ | ⋮ | ⋮ | ⋮ | ⋮ | ⋮ |

In the assumed example in Table 7, choice (1) derived by PMA did not match choice (4) derived by CCE. This is because PMA is based on permissible range expansion and CCE is based on preference order replacement. In CCE, the first preference order choice is meaningful and the high-rank choices are meaningful as a compromise range.

Therefore, if we rearrange the first place choice (4) in the CCE, the second place choice (6), and the PMA choice (1), we get "(4) no new nuclear power plants, but nuclear power plants can be restarted until alternative power generation methods are established, (6) nuclear power plants can be operated with the emphasis on safety, and (1) zero nuclear power plants by 2030." By synthesizing choices (4), (6), and (1), the choice "no new nuclear power plants, but restarting nuclear power plants is possible with the emphasis on safety until alternative power generation methods are established, to attain zero nuclear power plants by 2030" emerges. Therefore, choices (4), (6), and (1) do not conflict but suggest a sublated choice that combines them. Otherwise expressed, a consensus process that combines PMA and CCE allows for an eclectic mix of both outcomes to reach a sublation.

## 5    Discussion

While conventional discussion support technologies have difficulty in aggregating opinions, and GMCR, a conventional conflict resolution technology, is burdened with huge computational complexity (computational complexity: number of choices to the power of the number of participants), PMA (computational complexity: number of choices), a conflict resolution technology, and CCE (computational complexity: factorial of the number of choices), a newly proposed technology, can support facilitation by presenting agreeable proposals toward consensus of all members while reducing computational complexity.

In PMA, as shown in Table 8, a consensus is reached by selecting a choice that falls within the permissible ranges of all participants. However, as this consensusable choice minimizes the extent of the permissible range of all participants, the consensus it produces tends to be biased towards the majority.

In CCE, as shown in Tables 6 and 9, by selecting the choice with the smallest $Score$, a consensusable choice is derived with the fewest possible compromises for each participant and with a high degree of fairness. In Eq. 9, which was used to calculate the $Score$, the average $\mu$ and the $Standard\ deviation\ \sigma$ are added with the same weighting factor, but if it is desirable to reduce compromise and emphasize the majority, increase the factor of $\mu$, and to respect the minority and emphasize fairness, increase the factor of $\sigma$. Thus, the weighting of compromise and fairness can be varied depending on the social issue.

Note that, as shown in Tables 8 and 9, there may be PMA and CCE cases that cannot be reduced to one consensusable choice. For example, if either of them can derive one consensusable choice, then that choice should be preferred as the consensusable choice. If there are multiple consensusable choices on both sides and there is overlap, we can reduce the consensusable choices from the overlap. If there are multiple consensusable choices for both and there is no overlap, which is an unusual case, a new sublated choice must be derived based on the contents of both consensusable choices, as discussed at the end of Section 4.2.

A composite consensus-building process that combines PMA and CCE can provide consensusable choices that fall within the permissibility range of all participants or that balance the number of replacements (compromise) and fairness. If consensus is not reached by PMA, it can proceed to CCE, and if consensus is still not reached, it can facilitate the creation of a new sublated choice from both PMA's and CCE's consensusable choices.

## 6    Conclusions

We developed a new compromise choice exploration technology aimed at achieving consensus and provided a new composite consensus-building process that combines PMA and CCE. We conducted a trial experiment according to this process, and based on the results, we obtained the following findings.

- We confirmed that PMA would derive a choice that was within the permissible range of all participants that would lead to consensus. However, it was found that consensusable choices that were biased toward the majority tended to be derived because the choice that minimized the permissible range of all the participants was prioritized.

- We confirmed that CCE derives a consensusable choice for participants with the fewest possible compromises and with high fairness by deriving a preference order that minimizes the *Score*, which is the sum of the average $\mu$ and the *Standard deviation $\sigma$* of the number of replacements in the preference order. We also found that the weight coefficients of the average $\mu$ and *Standard deviation $\sigma$* can change the balance between the overall number of replacements (compromise) and fairness.
- The composite consensus-building process can provide a consensusable choice that emphasizes fairness, where CCE provides an equal degree of compromise for all if PMA does not lead to consensus. Even if consensus cannot be reached using both, a sublated choice can be obtained via SCC, which synthesizes the consensusable choices of PMA and CCE.

Based on this trial experiment, it was found that the priority and overlap of choices should be considered when it is not possible to reduce to a single consensusable choice in PMA or CCE. It was also necessary for CCE to change the weighting of the average number of replacements (compromise) and standard deviation (fairness) of the *Score* depending on the social issue. In addition, it would be desirable to work on technology to create a sublated choice in the future for cases where a consensusable choice cannot be found in either PMA or CCE.

The proposed consensus process consists of PMA and CCE, so in the future, it will be combined with an online discussion platform that has a series of functions such as proposal, discussion, facilitation, and decision. Although this trial experiment was conducted on a small scale, which represents a study limitation, we intend to conduct controlled experiments and fieldwork on a statistically significant scale targeting municipalities and local communities to put this method to practical use as a group decision-making method for solving social problems.

In the future, our proposed approach could be applied to a wide range of practical situations, from local issues in municipalities and communities to international issues such as environmental protection and human rights issues. It could also aid in the development of digital democracy [35] and platform cooperativism [36]. The former cites freedom and equality using digital technology, while the latter refers to joint ownership and the democratic governance of information platforms. We believe that the composite consensus-building process presented in this study will contribute to these movements.

**Acknowledgments:** The research for this study was conducted collaboratively between the Tokyo Institute of Technology and Hitachi Ltd. We thank Professor Takehiro Inohara of the Tokyo Institute of Technology for providing helpful suggestions about GMCR and PMA. We also received valuable advice from the Hitachi Kyoto University Laboratory of the Kyoto University Open Innovation Institute regarding suitable approaches to this research. To them, we express our deepest gratitude. There are no conflicts of interest to declare associated with this manuscript or study. The datasets generated and analyzed during the current study are available from the corresponding author upon reasonable request.